# The Single–Degenerate Scenario for Type Ia SNe in Cosmic Perspective

R. Canal[1], P. Ruiz-Lapuente[1,2], and A. Burkert[3]

[1]Department of Astronomy, University of Barcelona, Martí i Franqués 1, E–08028 Barcelona, Spain. E–mail: ramon@farcm0.am.ub.es, pilar@mizar.am.ub.es

[2]Max–Planck–Institut für Astrophysik, Karl–Schwarzschild–Strasse 1, D–85740 Garching, Federal Republic of Germany. E–mail: pilar@MPA–Garching.MPG.DE

[3]Max–Plank–Institut für Astronomie, Königstuhl 17, D–69117 Heidelberg, Federal Republic of Germany. E–mail: burkert@mpia–hd.mpg.de

## ABSTRACT

The occurrence and properties of Type Ia supernovae (SNe Ia) in single–degenerate binary systems (white dwarf [WD] + nondegenerate companion) is examined for galaxies of different types, and as a function of redshift. The rates and characteristics (peak luminosities, expansion velocities of the ejecta) expected from the explosion of mass–accreting WDs in symbiotic systems and "helium star cataclysmics" are found to be different from those arising in another class of candidate systems: cataclysmic–like (contact) systems (CLS), where a CO WD accretes hydrogen on a thermal time scale from a Roche–lobe filling main–sequence or subgiant companion. We derive the evolution of the SNe Ia rate and properties resulting from the thermonuclear explosion of sub–Chandrasekhar mass WDs in such systems when they detonate a helium layer accumulated from steady burning of hydrogen at the surface. A fraction of CLS are believed to form a subset of the observed luminous supersoft X–ray sources (SSS). Sub–Chandrasekhar explosions from CLS are disfavored in all types of galaxies at redshifts $z \gtrsim 1$. On the other hand, CLS where the WD succeeds to grow to the Chandrasekhar mass are more likely found in spiral galaxies, and absent from early–type galaxies. SNe Ia statistics could (if the uncertainties still involved are reduced) help to discriminate among proposed SNe Ia scenarios. The range of variation of the characteristics of SNe Ia in the CLS scenario should be narrower than in symbiotics. The predicted correlation between peak luminosity and velocity of the ejecta in SNe Ia coming from these systems is weak. For CLS, the distinction between the characteristics of SNe Ia respectively arising from sub–Chandrasekhar and from Chandrasekhar–mass explosions should be sharp, since all sub–Chandrasekhar explosions would be produced by low–mass WDs.

*Subject headings:* binaries: close — supernovae: general — X–rays: stars

## 1. Introduction

Correlations have been found between the peak luminosity of Type Ia supernovae (SNe Ia), the expansion velocity of their ejecta, and SNe Ia rates with the Hubble type of the parent galaxy: the luminosities, expansion velocities, and rates increase when going along the Hubble sequence from early–type to late–type galaxies (Filippenko 1989; van den Bergh & Pazder 1992; Branch & van den Bergh 1993; Maza et al. 1994; Della Valle & Livio 1994; Hamuy et al. 1994). The observed trends are those expected if CO white dwarfs (WDs) with masses lower than the Chandrasekhar mass would explode (Ruiz–Lapuente, Burkert, & Canal 1995, hereafter RBC95). Explosion of sub–Chandrasekhar CO WDs can be induced by detonation of a helium layer accumulated close to the surface either by direct accretion or from burning of accreted hydrogen (Livne & Glasner 1991; Ruiz–Lapuente et al. 1993; Woosley & Weaver 1994; Livne & Arnett 1995). Interest on explosions of sub–Chandrasekhar WDs has been stimulated by the discovery of subluminous SNe Ia (Filippenko et al. 1992; Leibundgut et al. 1993; Hamuy et al. 1995). Among possible systems producing SNe Ia by explosion of sub–Chandrasekhar WDs, symbiotic systems and "helium star cataclysmics" (Iben & Tutukov 1991; Limongi & Tornambe 1991) have been studied by RBC95. The results were compared with the CO + CO WD systems (double degenerate or DD systems) proposed by Iben & Tutukov (1984). It was shown that the right observed correlations might be obtained from symbiotic systems where the WD could explode below the Chandrasekhar mass, provided that the efficiency of retention of the accreted hydrogen by the WD were not too low. In this *Letter* we consider a further type of systems where sub–Chandrasekhar WD explosions might take place: cataclysmic–like (CLS) systems in which a CO WD accretes hydrogen on a thermal time scale from a Roche–lobe filling main–sequence or subgiant companion. Such systems were considered as possible SNe Ia progenitors (through Chandrasekhar–mass WD explosions) by Iben & Tutukov (1984).

More recently, van den Heuvel et al. (1992) have suggested steady nuclear burning on the surface of the mass–accreting WDs in those systems as a model for the luminous supersoft X–ray sources (SSS). SSS have been modelled by Rappaport, Di Stefano, & Smith (1994). They conclude that those systems might significantly contribute to the SNe Ia rate in our Galaxy. However, they only consider Chandrasekhar–mass explosions. We examine the possible contribution of sub–Chandrasekhar WD explosions in CLS to the SNe Ia rate and we compare it with those from symbiotic systems, He star cataclysmics, DD mergings, and also with that from Chandrasekhar–mass explosions in CLS. The time evolution of the rates for an instantaneous outburst of star formation, for an E0 galaxy, and for star formation rates which are typical for Sb and Sc galaxies is derived, together with the evolution of the explosion characteristics (average $^{56}$Ni yield and average explosion energy). We find that SNe Ia statistics up to large enough redshifts can help to discriminate among currently proposed SNe Ia scenarios.

## 2. Models and Results

To model the time evolution of the SNe Ia rates from CLS, we use the same Monte Carlo scenario code as in RBC95. The initial mass function (IMF) of the primary stars is that of Scalo (1986), and the distributions of initial mass ratios of the secondary to the primary ($q$) and of initial orbital periods ($P$) are taken from Duquennoy & Mayor (1991). This, together with conditions (1)–(3) below, constitutes our "standard" modeling. The history of the successful candidates to SNe Ia progenitors starts from binary systems with primaries with masses $1.8 M_\odot \lesssim M_1 \lesssim 8.0 M_\odot$ and secondaries with masses $M_2 \gtrsim 0.8 M_\odot$, in the period range $4 \ yr \lesssim P_{orb} \lesssim 11 \ yr$. They undergo an episode of common–envelope evolution when the primary is in the AGB stage of its evolution. The envelope of the primary is then lost, leaving a CO WD behind, while the orbital separation decreases. The primordial separation ($A_0$) is related to the final separation ($A_f$) through the common–envelope parameter $\alpha$, of order unity, which measures the efficiency of deposition of orbital energy into the ejection of the common envelope (see, for instance, Iben & Tutukov 1984). Numerical simulations (Livio & Soker 1988; Taam & Bodenheimer 1989) give values $\alpha \simeq 0.3 - 0.6$, but Yungelson et al. (1994) and Tutukov & Yungelson (1994) argue for higher values. We thus have initially adopted $\alpha = 1$. Adopting the value $\alpha = 0.3$ does not change the results for CLS but it affects the comparison with other types of systems. Further conditions are required in order to ensure that mass transfer will take place unstably on a thermal time scale, that no second common–envelope episode occurs, and that the mass–accretion rate is between the upper and the lower limit for stable thermonuclear burning of H into He. For the lower limit we first adopt the criterion derived by Iben (1982), and we will later discuss the influence of relaxing it. Roche–lobe filling by the secondary occurs due to its nuclear evolution.

The conditions for producing a sub–Chandrasekhar SNe Ia are: (1) the WD mass must be $M_{WD} \gtrsim 0.7 M_\odot$, (2) a minimum mass $\Delta M_{He} \gtrsim 0.1 M_\odot$ must accumulate at the surface of the WD, and (3) the mass–accretion rate must be in the range $10^{-9} M_\odot \ yr^{-1} \lesssim \dot{M} \lesssim 5 \times 10^{-8} M_\odot \ yr^{-1}$ (Nomoto 1982). The two first conditions have also been adopted by Yungelson et al. (1995) for production of SNe Ia by WDs belonging to symbiotic systems (see equally RBC95). For production of Chandrasekhar–mass explosions, conditions (1) and (2) change into: $M_{WD} + \Delta M = 1.4 M_\odot$. Condition (3) becomes: $5 \times 10^{-8} M_\odot \ yr^{-1} \lesssim \dot{M} \lesssim \dot{M}_{max}$, where $\dot{M}_{max}$ is the maximum accretion rate before the WD envelope expands to red–giant size. We first assume an instantaneous outburst of star formation and we follow the time evolution of the SNe Ia rate. That would be a first approximation to the case of an early–type galaxy. The result is shown in the first panel of Figure 1 (solid histogram): the first WDs start exploding at $t \simeq 2 - 3 \times 10^9 \ yr$ after outburst and the SNe Ia rate rises steeply, reaching its peak at $t \simeq 4 \times 10^9 \ yr$ to decrease again rapidly afterwards down to negligible values at $t \simeq 10$ Gyr. In contrast, SNe Ia from symbiotic systems would start already at $t \simeq 10^8$ $yr$, would reach a first peak at $t \simeq 2.5 \times 10^9$ $yr$, a second one at $t \simeq 6.5 \times 10^9$ $yr$, and decrease much less steeply afterwards (see RBC95). The average mass of the exploding WD in CLS steadily decreases with time, as it can be seen in the second panel of Figure 1. In that, the behaviour of symbiotics is similar. The amplitude of the variation, however, is larger in symbiotics than in CLS: it is $0.7 M_\odot \lesssim M_{WD} \lesssim 1.3 M_\odot$ in symbiotics and only $0.7 M_\odot \lesssim M_{WD} \lesssim 0.8 M_\odot$ in CLS. Due to the narrow range of WD masses, the average velocity of the ejecta (third panel of Figure 1) is almost constant ($M_{56_{Ni}}/M$ does only very slightly vary). Thus, in contrast with symbiotics, the predicted correlation of expansion velocity of the ejecta with peak luminosity in the CLS scenario is weak. To check the dependence on the value of the common–envelope parameter $\alpha$, we have repeated the calculation for $\alpha = 0.3$. The result is undistinguishable from that for $\alpha = 1$.

The lower limit derived by Iben (1982) for steady hydrogen burning in a mass–accreting WD ensures the absence of cycles along which the radius of the WD would significantly increase, thus possibly interfering in the accretion process. Accretion at lower rates (but still above the upper limit for producing nova–like outbursts) does not mean, however, that a helium mass large enough to produce a detonation cannot be accumulated. We have thus repeated the calculation taking $\dot{M} \gtrsim 10^{-9} M_\odot$ $yr^{-1}$ as the lower limit (which means a reduction by a factor four, approximately). The result is shown by the dot–dashed histogram in the first panel of Figure 1. Another important quantity is the upper limit on $\dot{M}$ ($\dot{M}_{upp}$) for He detonation to occur (condition 3 above). Increasing it by a factor two (from $5 \times 10^{-8}$ $M_\odot$ $yr^{-1}$ to $10^{-7}$ $M_\odot$ $yr^{-1}$) would not only increase the SNe Ia rate by a large factor but it would also advance the time of maximum rate after outburst by $\simeq 3$ Gyr (long–dashed line). As we will see (Figure 2 below), the latter would make the dependence of the rate on redshift for all types of galaxies, at $z \leq 1$, more similar to the case of Chandrasekhar–mass explosions (since $\dot{M}_{upp}$ would be closer to $\dot{M}_{max}$). In contrast, increasing $\Delta M_{He}$ (condition 2) also by a factor two (from 0.1 $M_\odot$ to 0.2 $M_\odot$) does not change the results. To test the sensitivity of the modeling to the IMF and to the $q$ and $P$ distributions, we have repeated the calculation (with "standard" values for conditions 1–3) for the IMF and $q$ and $A_0$ distributions adopted by Iben & Tutukov (1984) and Tutukov & Yungelson (1994). The comparison is made in the fourth panel of Figure 1. Although the peak rate approximately decreases by a factor three, its epoch does not change and the time evolution of the SNe Ia rates is roughly preserved. The dependence of rates and explosion characteristics on cosmological time and galaxy type is thus not significantly different from our "standard" case. Changing the IMF alone has a much smaller effect. A possible time–dependent IMF should thus include large deviations from the forms considered here to significantly alter the results.

Due to the precipitous decrease of the SNe Ia rate after reaching its maximum, the rate at the present time ($\sim$ 10 Gyr after the outburst) is very sensitive to the duration of the outburst. As a better approximation to the star–formation history in an early–type galaxy (E0) we have adopted the star–formation rates (SFRs) of Sandage (1986). For such type of galaxy, star formation spans over the first $\sim$ 1 Gyr of the galaxy's life. The result is shown in the bottom panel of Figure 2 as a function of redshift $z$ (for $z \leq 1$): the behaviour is qualitatively the same as for an instantaneous outburst, but now the SNe Ia rate at the present time would be $\nu_{SNeIa} \simeq 2 \times 10^{-5}$ $yr^{-1}$ for the most restrictive bound on $\dot{M}_{stable}$, and $\nu_{SNeIa} \simeq 4.6 \times 10^{-4}$ $yr^{-1}$ for the less restrictive one (in both cases for the "standard" $\dot{M}_{upp}$), in units of SNe $yr^{-1}$ per $10^{10}$ $M_\odot$ of parent galaxy, i.e. $\simeq$ SNe $yr^{-1}$ per $10^{10}$ $L_{bol}$ ($_\odot$). These rates, together with those from the explosion of sub–Chandrasekhar WDs in symbiotic systems and in helium cataclysmics, plus the DD merging rates and the rates of Chandrasekhar–mass explosions from CLS, are displayed in Table 1. Two different values are also given for the sub–Chandrasekhar SNe Ia rates from symbiotics: the larger one is from RBC95 and it would correspond to all symbiotics being able to accrete and retain enough hydrogen to eventually produce a helium detonation (clearly an upper limit). The lower value corresponds to the more restrictive conditions adopted by Yungelson et al. (1995). We see that, even assuming that DD mergings were 100% efficient in producing SNe Ia, sub–Chandrasekhar explosions from both CLS and symbiotics should dominate in early–type galaxies at the present time (especially if $\alpha < 1$). Moreover, a point too often ignored is that the efficiency of DD merging could actually be lower: off–center carbon ignition either during coalescence or by viscous heating in the accretion disk which forms afterwards has to be avoided and it appears difficult to prevent (see Mochkovitch & Livio 1989; Mochkovitch 1991). No sub–Chandrasekhar SNe Ia from CLS would be produced, however, at $z \gtrsim 0.8$ (for our "standard" choice of $\dot{M}_{upp}$), as it can be seen in the same panel. We also show the evolution of the SNe Ia rate from symbiotics for E0 galaxies. At low $z$, the slope of the curve for CLS is much steeper than that for symbiotics up to $z \simeq 0.35$. The SNe Ia at these redshifts are brighter in symbiotics than in CLS since they come from more massive WDs. The trend reverses afterwards and the first peak for symbiotics appears at a slightly larger redshift than the maximum rate from CLS. SNe Ia statistics might thus enable to discriminate between the two scenarios.

Adopting the SFRs proposed by Sandage (1986), we have calculated the time evolution of the SNe Ia rates from CLS for Sb and Sc galaxies both for Chandrasekhar and sub–Chandrasekhar explosions. It is shown, respectively, in the first and second panels of Figure 2. The rates at t = 10 Gyr are given in the fourth and fifth columns of Table 1, again for CLS, symbiotics, He cataclysmics, and DD. In late–type galaxies, the rates from any scenario (with one exception) are higher (by factors of 2–5) than in early–type galaxies. That is close to the trend found by Della Valle & Livio (1994). The predicted increase of the rate of SNe going towards late–type galaxies is a result of the process of continuous star formation in these galaxy types. Such an increase is smeared out if the rates are given in SNU (SNe per $10^{10}$ $L_B$ ($_\odot$) of parent galaxy luminosity and per 100 yr), since the ratio $M_{lum}/L_B$ decreases along the Hubble sequence. If they are referred to $L_{bol}$ or $L_K$, as in Della Valle & Livio (1994), the predicted trend should be observable. In order to compare with observations, rates both related to total mass of the galaxy and to $L_B$ are given using the ratios of the luminous mass of the galaxy (entering in the production of binaries) to its blue luminosity $M_{lum}/L_B$ from Blumenthal et al. (1984). We expect at the present time, in late–type galaxies, the SNe Ia rates from symbiotics and from He cataclysmics to be larger and those from DD merging to be similar to the rates from sub–Chandrasekhar CLS (for the most restrictive $\dot{M}_{stable}$, "standard" $\dot{M}_{upp}$, and if we assumed a 100% efficiency in SNe Ia production from DD merging), while CLS would give the dominant contribution to Chandrasekhar–mass explosions. For the less restrictive bound on $\dot{M}_{stable}$, only the maximal rates from symbiotics would be larger than those from CLS (both sub–Chandrasekhar and Chandrasekhar). Note that, in this case, the rates from sub–Chandrasekhar SNe Ia in CLS would be constant when going from early–type galaxies to late–type galaxies. Comparing the three panels of Figure 2 we see that if sub–Chandrasekhar WD explosions in CLS were the main progenitors of SNe Ia, the rates should increase more steeply with redshift in early–type than in late–type galaxies. The peak rates in E0, Sb, and Sc galaxies would correspond to $z \simeq 0.5$, $z \simeq 0.25$, and $z \simeq 0.2$, respectively, with our "standard" assumptions. On the other hand, no SNe Ia from Chandrasekhar–mass WD explosions in CLS should happen, in E0 galaxies, at $z \lesssim 0.35$.

We have examined the possible contribution of explosions of sub–Chandrasekhar WDs in CLS to the SNe Ia rate in different types of galaxies. We find that CLS might help to explain the observed trends of decreasing SNe Ia rates, decreasing luminosities, and lower velocities of the ejected material when going from late–type towards early–type galaxies. The correlation between peak luminosity and velocity of the ejecta should, in principle, be weak for this class of systems. Also, depending on the lower bound for $\dot{M}_{stable}$ (the lower limit for accumulating He on top of the WD from accretion of H), the rates might remain unchanged when going from early–type to late–type galaxies. The results are very sensitive to the value of $\dot{M}_{upp}$ (the upper limit for occurrence of He detonation): an increase of our "standard" value of $\dot{M}_{upp}$ by a factor two would make the behaviour of the sub–Chandrasekhar CLS SNe Ia rates much more similar to that of the Chandrasekhar ones. To recall previous conclusions (RBC95), DD merging, leading to Chandrasekhar–mass WD explosions, should be entirely absent from early–type galaxies if the common–envelope parameter $\alpha$ were significantly smaller than 1. The same would happen for Chandrasekhar–mass SNe Ia from CLS, and for SNe Ia from CO WD + He star systems (either Chandrasekhar or sub–Chandrasekhar), irrespective of the value of $\alpha$. That poses the question of which single–degenerate scenario would give a better account of the rates and characteristics of SNe Ia in early–type galaxies: symbiotics or CLS? If CLS were the sole progenitors of SNe Ia, there should be no explosions at $z \gtrsim 1.7$, even in E0 galaxies ($z \gtrsim 0.8$ for "standard" $\dot{M}_{upp}$). On the other hand, at lower redshifts, the SNe Ia rate should increase faster with redshift in early–type galaxies than in late–type ones. In contrast, SNe Ia from symbiotic systems would start at $z \simeq 2.5$ already, in E0 galaxies, and the increase of the rate with redshift, at low redshifts, would not be much faster in E0 than in Sb or Sc galaxies. SNe Ia from He cataclysmics would be absent from E0 galaxies at $z \lesssim 0.6$. Inclusion of CLS among the possible progenitors of SNe Ia in which the mass of the exploding WD were below the Chandrasekhar mass lends additional support to the idea that sub–Chandrasekhar WD explosions may give rise to a substantial fraction, at least, of SNe Ia, especially in early–type galaxies. The observed correlations of rates, luminosities, and expansion velocities with Hubble type of the parent galaxy would reflect this fact. The SNe Ia rates from CLS that we have derived should be close to the actual ones (for the right choice of $\dot{M}_{upp}$), since the selection criteria adopted for the progenitor systems do constrain mass–accretion rates by the WD to be in the appropriate range to produce helium detonations close to the surface (or to grow to the Chandrasekhar mass). The same is true for He cataclysmics. The SNe Ia rates from symbiotics should be close to the lower values in Table 1, which take into account the efficiency in the accretion and retention of hydrogen. *The distribution of SNe Ia properties expected from CLS is different to that expected from symbiotics: SNe Ia in CLS would be either Chandrasekhar–mass WD explosions (absent from early–type galaxies) or low–mass ($M_{WD} \lesssim 0.8 M_\odot$) sub–Chandrasekhar ones. In contrast, WD explosions in symbiotics would always be sub–Chandrasekhar but they would span a much broader mass range.* Observations should clarify whether the actual distribution of SNe Ia characteristics is either bimodal or continuous.

|  |  | | Galaxy | Type |
| --- | --- | --- | --- | --- |
| Scenario | $\alpha$ | E0 | Sb | Sc |
| CLS[a] | 1 | **2.0×10⁻⁵** | **1.1×10⁻⁴** | **1.1×10⁻⁴** |
|  |  | 0.02 | 0.03 | 0.02 |
| CLS[b] | 1 | **4.6×10⁻⁴** | **4.6×10⁻⁴** | **4.6×10⁻⁴** |
|  |  | 0.37 | 0.11 | 0.07 |
| Symbiotics[c] | 1 | **1.4×10⁻³** | **4.8×10⁻³** | **4.8×10⁻³** |
|  |  | 1.12 | 1.20 | 0.72 |
| Symbiotics[d] | 1 | **7.0×10⁻⁵** | **2.7×10⁻⁴** | **3.2×10⁻⁴** |
|  |  | 0.06 | 0.07 | 0.05 |
| He cataclys. | 1 | **0** | **2.3×10⁻⁴** | **4.6×10⁻⁴** |
|  |  | 0 | 0.06 | 0.07 |
| He cataclys. | 0.3 | **0** | **4.6×10⁻⁴** | **9.1×10⁻⁴** |
|  |  | 0 | 0.11 | 0.14 |
| DD[e] | 1 | **7.0×10⁻⁵** | **1.6×10⁻⁴** | **2.0×10⁻⁴** |
|  |  | 0.06 | 0.04 | 0.03 |
| DD[e] | 0.3 | **0** | **1.4×10⁻⁴** | **1.8×10⁻⁴** |
|  |  | 0 | 0.03 | 0.03 |
| CLS[e] | 1 | **0** | **4.6×10⁻⁴** | **4.6×10⁻⁴** |
|  |  | 0 | 0.11 | 0.07 |

Table 1: SNe Ia Rates at t = 10 Gyr[1]

---

[1] Boldface figures are rates per galaxy of $10^{10}\ M_\odot$. Roman figures are rates in SNU (thus refering to $L_B$ of the parent galaxy)

[a] More restrictive bound on $\dot{M}_{stable}$

[b] Less restrictive bound on $\dot{M}_{stable}$

[c] From RBC95

[d] From Yungelson et al. (1995)

[e] Chandrasekhar–mass explosions

---

## Figure Captions

*Figure 1.* Top panel: time evolution of the SNe Ia rates from explosion of sub–Chandrasekhar WDs in cataclysmic–like systems (CLS), after an instantaneous outburst of star formation. Second panel: time evolution of the average mass of the exploding WD (in $M_\odot$). Third panel: time evolution of the average velocity of the ejecta (in units of $10^4$ $km\ s^{-1}$). Dot–dashed histograms correspond to the less restrictive lower bound on $\dot{M}_{stable}$, and long–dashed ones to the higher value of $\dot{M}_{upp}$ (see text). In the top panel, the latter histogram is scaled down by a factor 2.5. In panels 2 and 3, the results for the symbiotic scenario are indicated by dotted lines. Bottom panel: the result of our "standard" modeling (solid histogram) is compared with that obtained for the IMF and $q$, $A_0$ distributions adopted by Iben & Tutukov (1984) and Tutukov & Yungelson (1994) (dotted histogram) (see text).

*Figure 2.* Top panel: The SNe Ia rate (in SNU) from CLS as a function of redshift $z$, for Sb galaxies. Continuous line corresponds to sub–Chandrasekhar WD explosions, for the "standard" values of $\dot{M}_{stable}$ and $\dot{M}_{upp}$ (see text). Long–dashed line, to the higher $\dot{M}_{upp}$, scaled down by a factor 10. Dotted line, to Chandrasekhar–mass explosions (scaled down by the same factor). Middle panel: Same, for Sc galaxies. Bottom panel: E0 galaxies. Solid line (A) is the "standard" sub–Chandrasekhar CLS case. Dot–dashed line (B) corresponds to the lower value of $\dot{M}_{stable}$. Long–dashed line (C), to the higher $\dot{M}_{upp}$. Dotted line (D) are Chandrasekhar–mass CLS explosions, and the short–dashed one (E) is the SNe Ia rate from symbiotics (scaled down by a factor 4). Transformation into redshift space is done as in RCB95.

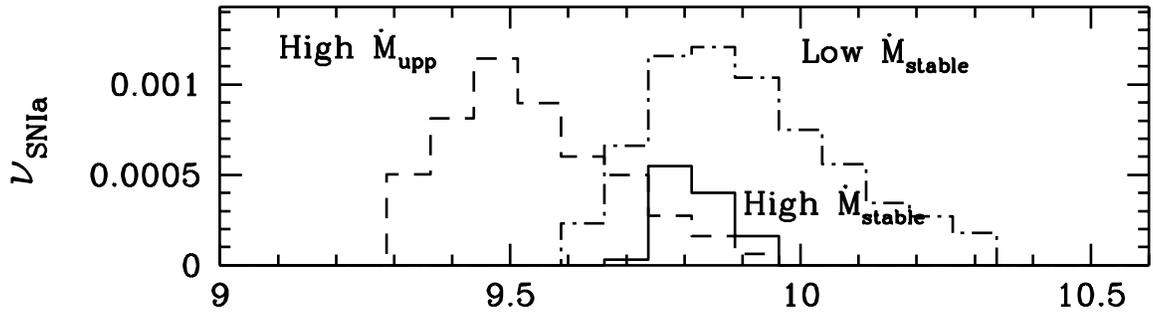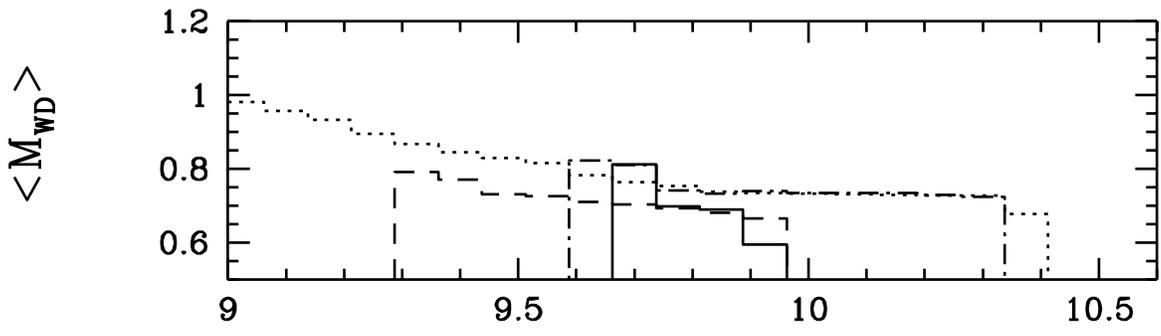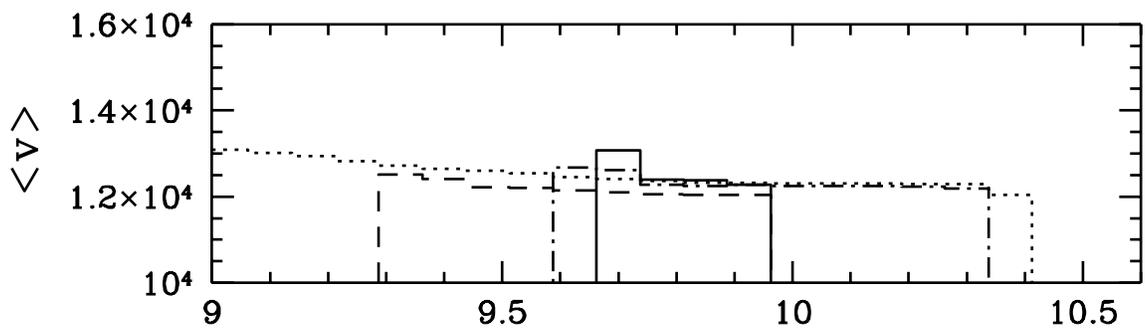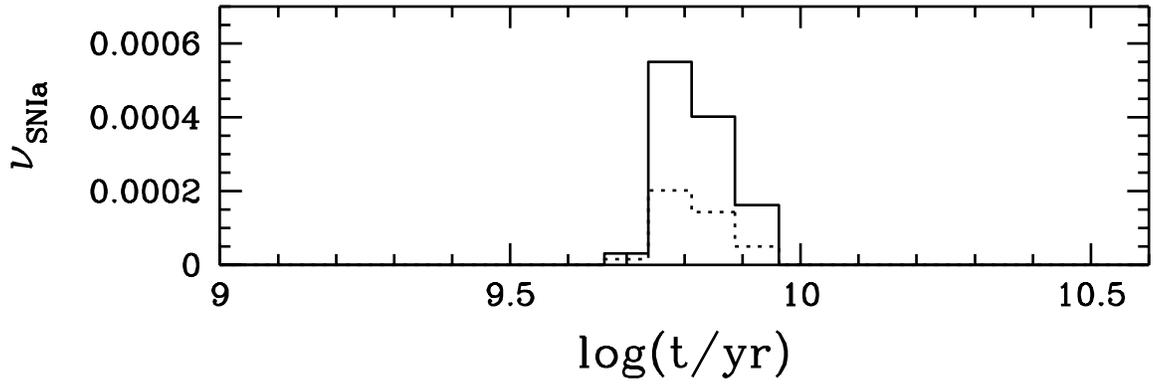

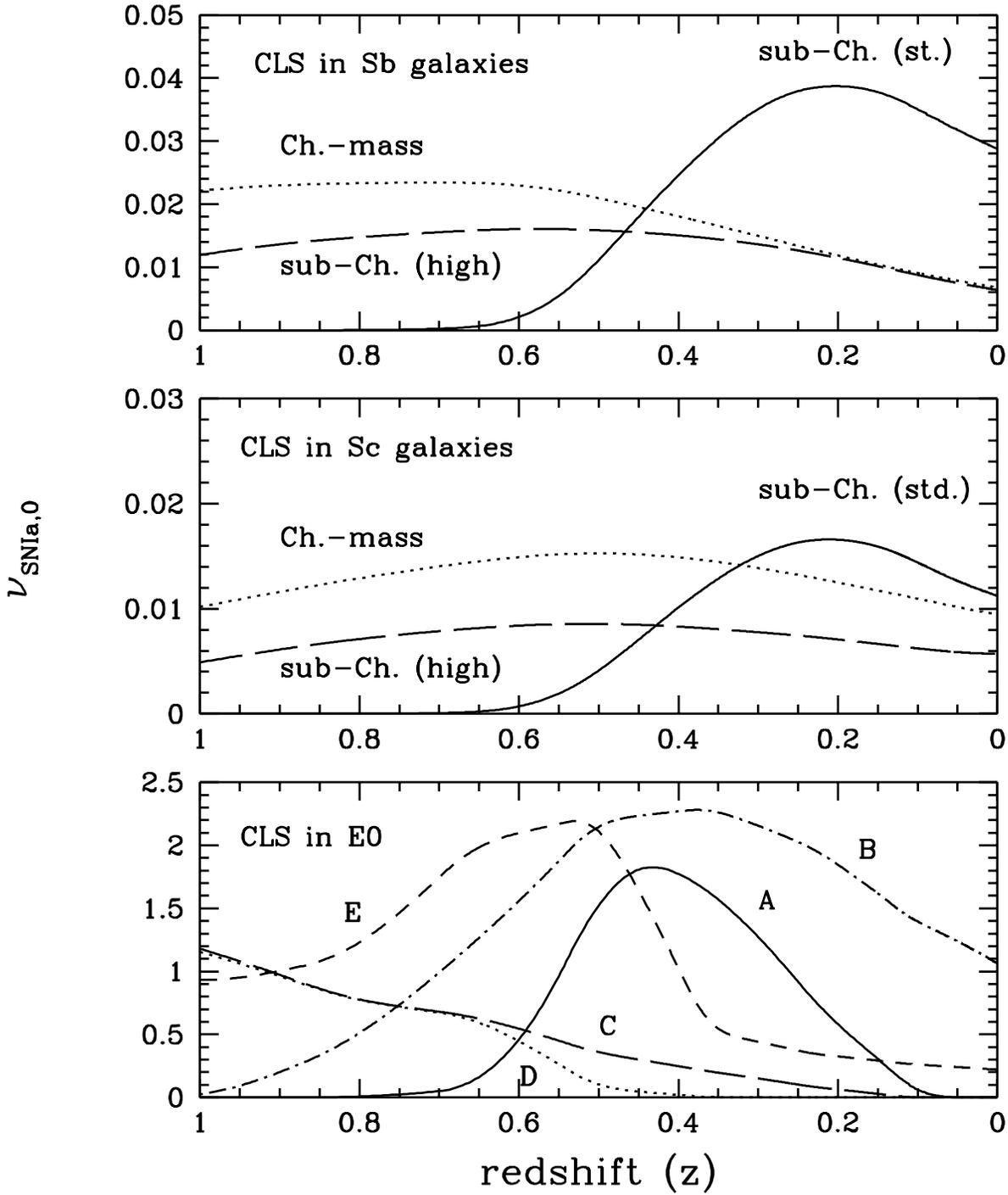